\documentstyle[11pt,epsf]{article}
\newtheorem{theorem}{Theorem}
\newtheorem{lemma}{Lemma}
\newcommand {\lexe} {\stackrel{\cdot} {\le}}
\newcommand {\gexe} {\stackrel{\cdot} {\ge}}
\newcommand {\dfn} {\stackrel{\Delta} {=}}
\newcommand {\exe} {\stackrel{\cdot} {=}}
\newcommand{\eqa}{\stackrel{\mbox{\tiny (a)}}{=}}
\newcommand{\eqb}{\stackrel{\mbox{\tiny (b)}}{=}}

\newcommand {\bx} {\mbox{\boldmath $x$}}
\newcommand {\by} {\mbox{\boldmath $y$}}

\newcommand {\bE} {\mbox{\boldmath $E$}}

\newcommand {\bR} {\mbox{\boldmath $R$}}

\newcommand {\bX} {\mbox{\boldmath $X$}}
\newcommand {\bY} {\mbox{\boldmath $Y$}}

\newcommand{\calA}{{\cal A}}
\newcommand{\calB}{{\cal B}}
\newcommand{\calC}{{\cal C}}

\newcommand{\calE}{{\cal E}}

\newcommand{\calI}{{\cal I}}

\newcommand{\calR}{{\cal R}}
\newcommand{\calS}{{\cal S}}
\newcommand{\calT}{{\cal T}}
\newcommand{\calU}{{\cal U}}
\newcommand{\calV}{{\cal V}}

\newcommand{\calX}{{\cal X}}
\newcommand{\calY}{{\cal Y}}

\newcommand {\tx} {\tilde{x}}
\newcommand {\ty} {\tilde{y}}
\newcommand {\tbx} {\tilde{\mbox{\boldmath $x$}}}
\newcommand {\hbx} {\hat{\mbox{\boldmath $x$}}}
\newcommand {\hbX} {\hat{\mbox{\boldmath $X$}}}
\newcommand {\tby} {\tilde{\mbox{\boldmath $y$}}}
\newcommand {\hby} {\hat{\mbox{\boldmath $y$}}}
\newcommand {\hbY} {\hat{\mbox{\boldmath $Y$}}}
\newcommand {\hH} {\hat{H}}
\newcommand {\hP} {\hat{P}}

\begin{document}
\thispagestyle{empty}
\title{Universal Decoding for Asynchronous Slepian-Wolf Encoding}
\author{Neri Merhav
}
\date{}
\maketitle

\begin{center}
The Andrew \& Erna Viterbi Faculty of Electrical Engineering\\
Technion - Israel Institute of Technology \\
Technion City, Haifa 32000, ISRAEL \\
E--mail: {\tt merhav@ee.technion.ac.il}\\
\end{center}
\vspace{1.5\baselineskip}
\setlength{\baselineskip}{1.5\baselineskip}

\begin{abstract}
We consider the problem of (almost) lossless source 
coding of two correlated memoryless sources
using separate encoders and a joint decoder, that is, Slepian-Wolf (S-W) coding.
In our setting, the encoding and decoding are
asynchronous, i.e., there is a certain relative delay between the two sources.
Neither the source parameters nor the relative delay are 
known to the encoders and the decoder. 
Since we assume that both encoders implement standard
random binning, which does not require such knowledge anyway, the focus 
of this work is on the decoder. Our main contribution is in proposing a universal 
decoder, that independent of the unknown source parameters and 
the relative delay, and at the same time, is asymptotically as good as the optimal 
maximum a posteriori probability (MAP) decoder in the sense of the 
random coding error exponent achieved.
Consequently, the achievable rate region is also the same as if the source parameters and
the delay were known to the decoder.\\

\end{abstract}

\newpage
\section{Introduction}

The problem of separate encodings and joint decoding of correlated sources, i.e., the well
known Slepian--Wolf (S-W) coding problem, has received a vast level of attention ever since
the landmark paper by Slepian and Wolf \cite{SW73} was published, nearly five decades ago.
Much less attention, however, was given to the asynchronous version of this problem, where
there is a relative delay between the two correlated sources, see, e.g.,
\cite{OO97}, \cite{TU20}, \cite{RU97}, \cite{STCW10}, \cite{Willems88}. 
The motivation for the asynchronous setting is thoroughly discussed in \cite{TU20}.
For memoryless
correlated sources, Willems \cite{Willems88} 
assumed that the relative delay is unknown to the encoders, but
known to the decoder, and proved that the achievable rate region is the same as in
synchronous S-W coding. Under similar assumptions, Rimoldi and Urbanke \cite{RU97}, as well
as Sun, Tian, Chen and Wong \cite{STCW10}, have proposed 
S-W data compression schemes that are
based on the notion of source splitting. In all these studies, it was assumed that the
decoder has the option to postpone the actual decoding until 
after having received all codewords associated with the data to be decoded. 
Such an assumption essentially neutralizes the negative effect of the relative delay
because the encoders and the decoder can still exploit the correlations between the two
sources. As explained, however, by Matsuta and Uyematsu in their recent paper 
\cite{TU20}, this setup might be
somewhat problematic, in practice, especially when the relative delay is very large.

The main result provided by Matsuta and Uyematsu in \cite{TU20} (see also \cite{TU13} and
\cite{TU15}) is 
a worst--case result in spirit. They assumed that: 
(i) the joint probability distribution, $P_{XY}$, of the
two corresponding correlated random variables, $X$ and $Y$, 
one from each source to be compressed,
is only known to belong to a subset $\calS$ of joint
probability distributions, (ii) 
the relative delay between the sources, $d$, is unknown, 
but known to be bounded between two
limits, and (iii) the absolute 
value of the relative delay, $|d|$, is allowed
to scale linearly with $n$, and the ratio $|d|/n$ 
tends to a constant, $\delta\in [0,1]$, as $n\to\infty$, and $\delta$ is only
known to be upper bounded by a given number $\Delta$.
They have proved a coding theorem asserting that the achievability 
rate region is as follows.
A rate--pair $(R_x,R_y)$ is achievable if and only if it satisfies 
the following three inequalities at the same time:
\begin{eqnarray}
\label{rx}
R_x\ge&
\sup_{P_{XY}\in\calS}[H(X|Y)+\Delta I(X;Y)]\\
\label{ry}
R_y\ge&\sup_{P_{XY}\in\calS}[H(Y|X)+\Delta I(X;Y)]\\
\label{rxry}
R_x+R_y\ge&\sup_{P_{XY}\in\calS}[H(X,Y)+\Delta I(X;Y)],
\end{eqnarray}
This result of \cite{TU20} is very interesting, but it is also extremely pessimistic.
It is overly pessimistic, not only because it is worst--case result, 
but even more importantly, because
the above three suprema can potentially be achieved by {\it three different
sources}, in general. Thus, for a single, given underlying joint source, 
$P_{XY}\in\calS$ (with relative delay, $\delta n$), 
as bad as it may be, at least one of the
above three inequalities can be improved, in general.
Moreover, if $\calS$ happens to be the entire simplex of probability 
distributions $\{P_{XY}\}$ over the given alphabets, $\calX$ and $\calY$
(which is a very realistic special case), 
these suprema are given by $\log|\calX|$, $\log|\calY|$,
and $\log|\calX|+\log|\calY|$, respectively, rendering this coding theorem 
an uninteresting triviality, as it allows no compression at all. 
The fact of the matter is, however, that at least the
weakness concerning the three different 
achievers of the suprema in (\ref{rx})--(\ref{rxry}) 
can be handled rather easily. Upon a careful inspection of the proof of
the converse part in \cite{TU20}, one readily concludes that it actually supports
an assertion that the achievable rate region is included in the following set:
\begin{eqnarray}
\bigcap_{P_{XY}\in\calS}\bigg\{(R_x,R_y):~R_x&\ge&H(X|Y)+\Delta I(X;Y),\nonumber\\
R_y&\ge&
H(Y|X)+\Delta I(X;Y),\nonumber\\
R_x+R_y&\ge&H(X,Y)+\Delta I(X;Y)\bigg\}.
\end{eqnarray}
Similar comments apply to the analysis of the error probability in \cite{TU20}, which
is a pessimistic analysis, carried out for the worst
source in $\calS$ and over all possible relative delay values, 
rather than the actual error probability associated with a given underlying
source.

In this paper, we tackle the problem in a different manner.
Instead of a worst--case approach, our approach is the
following: for a given rate pair $(R_x,R_y)$, even if we knew the 
source and the relative delay, we could have
handled only sources, $\{P_{XY}\}$, that satisfy $H(X|Y)+\delta
I(X;Y)\le R_x$, $H(Y|X)+\delta I(X;Y)\le R_y$ and $H(X,Y)+\delta I(X;Y)\le
R_x+R_y$ ($\delta$ being the actual normalized relative delay). 
Now, the S-W encoders are always simple random--binning encoders, regardless of
the source parameters, so every uncertainty, that is 
associated with the source parameters and the
relative delay, is confronted, and therefore must be handled, by the decoder.
Owing to the earlier results on the 
minimum--entropy universal decoder for the S-W encoding system
(see, e.g., \cite{CHJL08}, 
\cite{Csiszar80}, \cite{Csiszar82}, \cite{CK80}, 
\cite[Exercise 3.1.6]{CK81}, \cite{Draper04}, 
\cite{Kieffer80}, \cite{p183}, \cite{OH94}, \cite{SBB05}),
it is natural to set the goal of seeking a universal decoder 
that is asymptotically as good as the optimal 
maximum a posterior (MAP) decoder for the given
source, in the sense that the random coding error exponent is the same, and hence so
is the achievable rate region. In other words, 
unlike in previous works on universal S-W decoding, here
universality is sought, not only with respect to (w.r.t.) the source distribution,
$P_{XY}$, but also w.r.t.\ the unknown relative delay between the two parts of the source.
Although it is natural to think of the relative delay 
as of yet another unknown parameter associated with the underlying source, it will be
interesting to see that in our proposed 
universal decoder, the unknown delay will be handled
differently than the other unknown parameters. We will elaborate on this point later on.

Our main contributions, in this work, are the following:
\begin{enumerate}
\item We propose a universal decoder that allows uncertainty, not only regarding the
source parameters, but also the relative delay.
\item We prove that our universal decoder achieves the same error exponent as the
optimal MAP decoder that is cognizant of both the source parameters and the relative delay.
This will be done by showing that our upper bound on the error probability 
of the universal decoder is of the same exponential order as a lower bound on the
error probability of the MAP decoder.
\item We provide the Lagrange--dual form of the resulting error exponent, and thereby
characterize the achievable rate region for achieving a prescribed random coding error
exponent, $E$.
\item We provide an outline for a possible extension to sources with memory.
\end{enumerate}

The outline of the remaining part of this paper is as follows.
In Section \ref{notation}, we establish notation conventions.
In Section \ref{pf}, we formulate the problem and spell out the
objectives of this work. In Section \ref{mr}, we present the main theorem and
discuss it. Finally, in Section \ref{proof}, we prove this theorem.

\section{Notation Conventions}
\label{notation}

Throughout the paper, random variables will be denoted by capital
letters, specific values they may take will be denoted by the
corresponding lower case letters, and their alphabets
will be denoted by calligraphic letters. Random
vectors and their realizations will be denoted,
respectively, by capital letters and the corresponding lower case letters,
both in the bold face font. Their alphabets will be superscripted by their
dimensions. For example, the random vector $\bX=(X_1,\ldots,X_n)$, ($n$ --
positive integer) may take a specific vector value 
$\bx=(x_1,\ldots,x_n)$ in $\calX^n$, the $n$--th order Cartesian power of $\calX$, which 
is the alphabet of each component of this vector. Segments of vector components will be
denoted by subscripts and superscripts, for example,
$x_i^j$, $i< j$, will designate
the segment $(x_i,x_{i+1},\ldots,x_j)$. When $i=1$, the subscript will be omitted and
therefore the notation will be $x^j$. By convention, when $i>j$, $x_i^j$ will be understood
to be the empty string, whose probability is formally defined to be unity.
Sources and channels will be denoted by the letter $P$ or $Q$,
subscripted by the names of the relevant random variables/vectors and their
conditionings, if applicable, following the standard notation conventions,
e.g., $Q_X$, $P_{Y|X}$, and so on. When there is no room for ambiguity, these
subscripts will be omitted.
The probability of an event $\calE$ will be denoted by $\mbox{Pr}\{\calE\}$,
and the expectation
operator with respect to (w.r.t.) a probability distribution $P$ will be
denoted by
$\bE_P\{\cdot\}$. Again, the subscript will be omitted if the underlying
probability distribution is clear from the context.
The entropy of a random variable (RV) $X$ with a generic distribution $Q$ will be denoted by
$H_Q(X)$. Similarly, other information measures will be denoted using the customary notation,
subscripted by the name of the underlying distribution $Q$. For example, for a pair of
RVs, $(X,Y)$, distributed according to $Q_{XY}$ (or $Q$, for short), $H_Q(X,Y)$,
$H_Q(X|Y)$ and $I_Q(X;Y)$ will denote the joint entropy, the 
conditional entropy of $X$ given $Y$, and the mutual information, respectively.
For two
positive sequences $a_n$ and $b_n$, the notation $a_n\exe b_n$ will
stand for equality in the exponential scale, that is,
$\lim_{n\to\infty}\frac{1}{n}\log \frac{a_n}{b_n}=0$. Similarly,
$a_n\lexe b_n$ means that
$\limsup_{n\to\infty}\frac{1}{n}\log \frac{a_n}{b_n}\le 0$, and so on.
The indicator function
of an event $\calE$ will be denoted by $\calI\{E\}$. The notation $[x]_+$
will stand for $\max\{0,x\}$.

The empirical distribution of a sequence $\bx\in\calX^n$, which will be
denoted by $\hat{P}_{\bx}$, is the vector of relative frequencies,
$\hat{P}_{\bx}(x)$,
of each symbol $x\in\calX$ in $\bx$.
The type class of $\bx\in\calX^n$, denoted $\calT(\bx)$, is the set of all
vectors $\bx'$
with $\hat{P}_{\bx'}=\hat{P}_{\bx}$. When we wish to emphasize the
dependence of the type class on the empirical distribution $\hat{P}$, we
will denote it by
$\calT(\hat{P})$, with a slight abuse of notation. 
Information measures associated with empirical distributions
will be denoted with `hats' and will include the
names of the vectors from which they are induced by parentheses.
For example, the empirical entropy of $\bx$, which is the entropy associated with
$\hat{P}_{\bx}$, will be denoted by
$\hat{H}(\bx)$. An alternative notation, following the conventions
described in the previous paragraph, is $H(\hP_{\bx})$.
Similar conventions will apply to the joint empirical
distribution, the joint type class, the conditional empirical distributions
and the conditional type classes associated with pairs of
sequences of length $n$.
Accordingly, $\hP_{\bx\by}$ will be the joint empirical
distribution of $(\bx,\by)=\{(x_i,y_i)\}_{i=1}^n$,
$\calT(\bx,\by)$ or $\calT(\hP_{\bx\by})$ will denote
the joint type class of $(\bx,\by)$, $\calT(\bx|\by)$ will stand for
the conditional type class of $\bx$ given
$\by$, $\hH(\bx,\by)$ will designate the empirical joint entropy of $\bx$
and $\by$,
$\hH(\bx|\by)$ will be the empirical conditional entropy,
and so on.
Clearly, empirical information measures can be calculated, not only from the
full vectors, but also from partial segments, like $x_i^j$ and $y_i^j$. In this case,
$x_i^j$ and $y_i^j$ will replace $\bx$ and $\by$ in the above notations.

\section{Problem Formulation}
\label{pf}

Let $\{(X_i,Y_i)\}$ be a pair of correlated 
discrete memoryless sources (DMSs) with a relative delay of $d$
time units, that is, $(X_i,Y_{i+d})$ are jointly distributed according to
a certain probability distribution, $P_{XY}$ for every $i$, but the
various pairs are mutually independent. In other words, the random vectors 
$Z_i=(X_i,Y_{i+d})$ are
i.i.d.\ for different values of $i$. Neither $P_{XY}$ and $d$ are known to the
encoders and decoder.

Similarly as in \cite{TU20}, the two separate encoders that compress $\{X_i\}$ and
$\{Y_i\}$ both operate on successive blocks of length $n$, without any attempt to align
them, because $d$ is unknown and it may be arbitrarily large. These encoders are ordinary
S-W encoders at rates $R_x$ and $R_y$, respectively. In other words, 
each member $\bx$ (resp.\ $\by$) of
$\calX^n$ (resp.\ $\calY^n$) is mapped into a bin $f(\bx)\in\{1,2,\ldots,
2^{nR_x}\}$ (resp.\ $g(\by)\in\{1,2,\ldots,2^{nR_y}\}$), 
which is selected independently at random
for every $n$--vector in the respective source space. As always, the 
randomly selected mappings of both encoders are revealed to the decoder.

As already mentioned, both $P_{XY}$ and $d$ are unknown,
but without essential loss of generality,
it may be assumed that $0\le d\le n$. For any $d\ge n$,
the respective blocks concurrently encoded, $\bX$ and $\bY$,
are statistically independent, and so,
all values of $d$, from $n$ and beyond, are actually equivalent from the viewpoints of
the encoders and the decoder.
The lower limit, $d\ge 0$, is
assumed for convenience only. Negative values of $d$ correspond to switching the roles
of the two sources in the forthcoming results and discussions
(see also \cite{TU20}).\footnote{We could have allowed negative values of $d$ in the
formal problem setup to begin with, but this would make the notation more cumbersome.}
Our asymptotic regime will be defined as described in the Introduction: as $n\to\infty$,
the relative delay $d$ will be asymptotically proportional to $n$, i.e., the
ratio $d/n$ tends to a limit, $\delta$. In view of the above discussion, $\delta$ can
be assumed to take on values in the interval $[0,1]$.

The decoder receives the bin indices, $f(\bx)$ and $g(\by)$, 
of the compressed vectors, $\bx$ and $\by$, respectively, and it outputs a pair of estimates,
$(\hbx,\hby)\in\calX^n\times\calY^n$. The average probability of error is defined
as
\begin{equation}
\bar{P}_{\mbox{\tiny e}}\dfn\mbox{Pr}\{(\hbX,\hbY)\ne(\bX,\bY)\},
\end{equation}
where both the randomness of $(\bX,\bY)$ and the randomness of the encoder mappings
are taken into account. The respective error exponent is defined as
\begin{equation}
E(R_x,R_y)=\lim_{n\to\infty}\left[-\frac{\log \bar{P}_{\mbox{\tiny e}}}{n}\right],
\end{equation}
provided that the limit exists.

The optimal MAP decoder, that is cognizant of both $P_{XY}$ and $d$, is given by
\begin{eqnarray}
(\hbx,\hby)&=&\mbox{arg}\max_{\{(\bx',\by'):~f(\bx')=f(\bx),~
g(\by')=g(\by)\}}P_d(\bx',\by')\\
&=&\mbox{arg}\min_{\{(\bx',\by'):~f(\bx')=f(\bx),~g(\by')=g(\by)\}}
\left\{-\log P_d(\bx',\by')\right\},
\end{eqnarray}
where
\begin{equation}
P_d(\bx,\by)=P(y_1^d)\cdot P(x_1^{n-d},y_{d+1}^n)\cdot P(x_{n-d+1}^n),
\end{equation}
where all three factors admit product forms,
\begin{eqnarray}
P(y_1^d)&=&\prod_{i=1}^dP_Y(y_i),\\
P(x_1^{n-d},y_{d+1}^n)&=&\prod_{i=1}^{n-d}P_{XY}(x_i,y_{i+d}),\\
P(x_{n-d+1}^n)&=&\prod_{i=n-d+1}^nP_X(x_i).
\end{eqnarray}
The average probability of error, associated with the MAP decoder, will be denoted
by $\bar{P}_{\mbox{\tiny e},\star}$ and its error exponent will be denoted
by $E_\star(R_x,R_y)$.
A general metric decoder is of the form
\begin{equation}
\label{gendec}
(\hbx,\hby)=\mbox{arg}\min_{\{(\bx',\by'):~f(\bx')=f(\bx),~g(\by')=g(\by)\}}q(\bx',\by'),
\end{equation}
where the function $q$ will be referred to as the {\it decoding metric}. The average error probability of the decoder that is based on the metric $q$, will be denoted by $P_{\mbox{\tiny e},q}$,
and its error exponent (if existent) will be denoted by $E_q(R_x,R_y)$.

In this paper, we propose a universal decoding metric $q$, that is independent of the
unknown $P_{XY}$ and $d$, yet its error exponent, $E_q(R_x,R_y)$, coincides with $E_*
(R_x,R_y)$, and hence it is asymptotically optimal in the random coding error--exponent sense.

\section{Main Result}
\label{mr}

We define the following functions for $0\le k\le n$:
\begin{eqnarray}
u_k(\bx,\by)&=&
k\hH(y_1^k)+(n-k)\hH(x_1^{n-k},y_{k+1}^n)+k\hH(x_{n-k+1}^n),\\
v_k(\bx,\by)&=&
(n-k)\hH(x_1^{n-k}|y_{k+1}^n)+k\hH(x_{n-k+1}^n),\\
w_k(\bx,\by)&=&
k\hH(y_1^k)+(n-k)\hH(y_{k+1}^n|x_1^{n-k}),\\
q_k(\bx,\by)&=&\max\{u_k(\bx,\by)-n(R_x+R_y),v_k(\bx,\by)-nR_x,w_k(\bx,\by)-nR_y\},
\end{eqnarray}
and finally, the universal decoding metric, $q$, is defined as
\begin{equation}
\label{me}
q(\bx,\by)=\min_{0\le k\le n}q_k(\bx,\by).
\end{equation}
If the relative delay, $d$, is allowed to take on also negative values, i.e., $-n\le d\le n$,
then the minimum in (\ref{me}) should be extended to $-n\le k\le n$, where for
$k < 0$, $u_k$, $v_k$, $w_k$, and $q_k$ are defined exactly as above, except that the
roles of $\bx$ and $\by$ are interchanged (that is, $\bx$ will be shifted 
$|k|$ positions to the right relative to $\by$, instead of the above shift,
which is the opposite).
For a pair of finite--alphabet RVs, $(X,Y)\sim P_{XY}$, let as define the R\'enyi
entropies of order $\theta > 0$ as
\begin{eqnarray}
H_\theta(X)&=&\frac{1}{1-\theta}\log\left\{\sum_{x\in\calX}[P_X(x)]^\theta\right\}\\
H_\theta(Y)&=&\frac{1}{1-\theta}\log\left\{\sum_{y\in\calY}[P_Y(y)]^\theta\right\}\\
H_\theta(X,Y)&=&\frac{1}{1-\theta}\log\left\{\sum_{(x,y)\in\calX\times\calY}[P_{XY}(x,y)]^\theta\right\}\\
H_\theta(X|Y)&=&\frac{\theta}{1-\theta}\log\left\{\sum_{y\in\calY}
\left[\sum_{x\in\calX}[P_{XY}(x,y)]^\theta\right]^{1/\theta}\right\}\\
H_\theta(Y|X)&=&\frac{\theta}{1-\theta}\log\left\{\sum_{x\in\calX}
\left[\sum_{y\in\calY}[P_{XY}(x,y)]^\theta\right]^{1/\theta}\right\}.
\end{eqnarray}
For $\theta\to 1$, these quantities tend to the respective Shannon entropies.

Our main result is the following.
\begin{theorem}
Under the assumptions formalized in Section \ref{pf}, the following is true:
\begin{enumerate}
\item[(a)] The error exponents, $E_*(R_x,R_y)$ and $E_q(R_x,R_y)$, both
exist.
\item[(b)] $E_q(R_x,R_y)=E_*(R_x,R_y)=\min
\{E_{x|y}(R_x),E_{y|x}(R_y),E_{xy}(R_x,R_y)\}$, where
\begin{eqnarray}
E_{x|y}(R_x)&=&\max_{0\le\rho\le 1}\rho\cdot\left[R_x-\delta H_{1/(1+\rho)}(X)-
(1-\delta)H_{1/(1+\rho)}(X|Y)\right]\\
E_{y|x}(R_y)&=&\max_{0\le\rho\le 1}\rho\cdot\left[R_y-\delta H_{1/(1+\rho)}(Y)-
(1-\delta)H_{1/(1+\rho)}(Y|X)\right]\\
E_{xy}(R_x,R_y)&=&\max_{0\le\rho\le 1} \rho\cdot\bigg[R_x+R_y-
\delta H_{1/(1+\rho)}(X)-\nonumber\\
& &\delta H_{1/(1+\rho)}(Y)-(1-\delta)H_{1/(1+\rho)}(X,Y)\bigg].
\end{eqnarray}
\end{enumerate}
\end{theorem}

\noindent
{\bf Discussion.} The remaining part of this section is devoted to a
discussion on Theorem 1 and its significance.

Since the error exponents were defined under the condition
that the certain limits exist, part (a) of the theorem establishes the basic
fact that they indeed exist. Part (b) is more quantitative: it tells that the
error exponents of the universal decoder and the MAP decoder are equal, thus
rendering the universal decoder asymptotically optimal in the error
exponent sense. Finally, part (b) provides also an exact single--letter
expression of this error exponent, using a Gallager--style formula. Here,
unlike in the synchronous case (of $\delta=0$), we also see unconditional
R\'enyi entropies (weighted by $\delta$), which correspond to the compression
of the segments,
$y_1^d$ and $x_{n-d+1}^n$, that are independent of each other and of all
other pieces of data within the block, and hence no correlations can be
exploited when compressing them. If $d$ is fixed (or grows sub--linearly with $n$), the
relative weight of these segments is asymptotically negligible, and there is
no asymptotic loss compared to the synchronous case. The error exponent is given by
the minimum among three error exponents: $E_{x|y}(R_x)$ corresponds to errors in
the decoding of $\bx$ while $\by$ is decoded correctly, $E_{y|x}(R_y)$ designates the
opposite type of error, and finally, $E_{xy}(R_x,R_y)$ stands for erroneous decoding of
both $\bx$ and $\by$. The smallest of all three dominates the overall error exponent.

The above relation between the error exponent and the coding rates can be essentially inverted, 
in order to answer the following question: what is the achievable rate region,
$\calR(E)$, for
achieving an error exponent at least as large as a prescribed value, $E$?
Using the above error exponent formula, the answer is readily
found\footnote{See the last part of Subsection \ref{formula}.} to be the
following.
\begin{equation}
\calR(E)=\{(R_x,R_y):~R_x\ge \bR_x(E),~R_y\ge\bR_y(E),~R_x+R_y\ge\bR_{xy}(E)\},
\end{equation}
where
\begin{eqnarray}
\bR_x(E)&=&\inf_{s\ge 1}\left[sE+\delta
H_{s/(1+s)}(X)+(1-\delta)H_{s/(1+s)}(X|Y)\right]\\
\bR_y(E)&=&\inf_{s\ge 1}\left[sE+\delta
H_{s/(1+s)}(Y)+(1-\delta)H_{s/(1+s)}(Y|X)\right]\\
\bR_{xy}(E)&=&\inf_{s\ge 1}\left[sE+\delta
H_{s/(1+s)}(X)+\delta H_{s/(1+s)}(Y)+(1-\delta)H_{s/(1+s)}(X,Y)\right].
\end{eqnarray}
For $E\to 0$, which means a vanishing error probability, however slowly, 
the infima are approached by $s\to\infty$, which yield
\begin{eqnarray}
\bR_x(0)&=&\delta H(X)+(1-\delta)H(X|Y)=H(X|Y)+\delta I(X;Y)\\
\bR_y(0)&=&\delta H(Y)+(1-\delta)H(Y|X)=H(Y|X)+\delta I(X;Y)\\
\bR_{xy}(0)&=&\delta H(X)+\delta H(Y)+(1-\delta)H(X,Y)=H(X,Y)+\delta I(X;Y),
\end{eqnarray}
as expected in view of the results of \cite{TU20}.

In order to try to understand the decoding metric (\ref{me}), consider the following
observations. This decoding metric is 
given by the maximum of three different metrics, which
are all in the spirit of the minimum entropy 
(ME) universal decoding metric,\footnote{The above defined function,
$u_k(\bx,\by)$, was mentioned also in \cite[Section V, second paragraph]{TU20}
as a possible decoding metric, but it was not the decoding metric actually analyzed there, because the
authors argued that it cannot be analyzed by the standard method of types.} but modified to
address the dependence structure at hand. Each
one of these metrics is `responsible' to handle a different type of error:
$u_k(\bx,\by)-n(R_x+R_y)$ is
associated with errors in decoding both $\bx$ and $\by$, $v_k(\bx,\by)-nR_x$ is for errors
in $\bx$ only, while $\by$ is decoded correctly, and finally,
$w_k(\bx,\by)-nR_y$ is meant for the
opposite case, of decoding error in $\by$ only. The maximum of all three
metrics is meant to handle
all three types of error at the same time. Every value of $k$ corresponds to
a certain hypothesis concerning the relative delay. 
Note that this decoding metric is different from the one in \cite{OH94}, which relies
on an encoding scheme that provides pointers to the type classes of $\bx$ and $\by$,
in addition to their bin indices.

Another observation is regarding the special stature of the relative delay
parameter, $d$. 
On the face of it, it is natural to view $d$ as yet another unknown parameter of the
source, in addition to the other unknown parameters -- those associated with the joint distribution, $P_{XY}$.
If $d$ was known, and only $P_{XY}$ was unknown, we could have interpreted the
empirical entropies in $u_k$, $v_k$ and $w_k$ (actually, with $k=d$) as negative logarithms of the
maximum likelihood (ML) values of the various segments, 
or equivalently, as the minima of the negative log--likelihood values.
For example,
$(n-k)\hH(x_1^{n-d},y_{d+1}^n)=\min_{P_{XY}}[-\log P(x_1^{n-d},y_{d+1}^n)]$,
$(n-k)\hH(x_1^{n-d}|y_{d+1}^n)=\min_{P_{XY}}[-\log P(x_1^{n-d}|y_{d+1}^n)]$, and
so on. In other words, the minima over $\{P_{XY}\}$ are taken {\it before} (i.e., more
internally to) the maximum over the three metrics. By contrast, the minimum
over the hypothesized relative delay, $k$, is taken {\it after} (i.e.,
externally to) the maximum over the three metrics. Attempts were made to prove
that minimum over $k$ and the maximum among the three metrics can be commuted,
but to no avail. Therefore, this point seems to be non--trivial.

Finally, it is in order to say a few words concerning sources with memory.
Consider the case where $Z_i=(X_i,Y_{i+d})$ is a first--order Markov source.
In this case, the techniques of \cite[Subsection 5.1]{p183}, suggest that one can prove the
universal asymptotic optimality of a similar universal decoder, where
$k\hH(y_1^k)$, $k\hH(x_{n-k+1}^n)$, $(n-k)\hH(x_{n-k+1}^n|y_{k+1}^n)$,
$(n-k)\hH(y_{k+1}^n|x_{n-k+1}^n)$, and $(n-k)\hH(x_{n-k+1}^n,y_{k+1}^n)$ are
replaced by the respective length functions associated with the Lempel--Ziv
algorithm (LZ78) \cite{ZL78} and the conditional LZ78 algorithm \cite{Ziv85},
$\mbox{LZ}(y_1^k)$, $\mbox{LZ}(x_{n-k+1}^n)$,
$\mbox{LZ}(x_{n-k+1}^n|y_{k+1}^n)$, $\mbox{LZ}(y_{k+1}^n|x_{n-k+1}^n)$, and
$\mbox{LZ}(x_{n-k+1}^n,y_{k+1}^n)$. 
It should be noted that $y_1^k$ and $x_{n-k+1}^n$ are
not realizations of a Markov sequences, but they are realizations of a
hidden-Markov process, as their correlated counterparts are not available. Nonetheless,
hidden Markov sources can still be accommodated in this framework
(see, e.g., \cite{p191}).

\section{Proof of Theorem 1}
\label{proof}

The proof is based on a simple sandwich argument: we first derive an upper
bound to the average error probability of the universal decoder that is based on $q$,
and then a lower bound to the error probability of the MAP decoder. Both
bounds turn out to be of the same exponential order. On the other hand, since the MAP decoder cannot be
worse than the universal decoder, this exponential order must be exact for
both decoders and its single--letter expression is easily
derived using the method of types. This will establish both part (a) of Theorem 1 and the first
equality in part (b). The second equality in part (b) will be obtained by deriving the
Lagrange--dual of the original single--letter formula.

\subsection{Upper Bound on the Error Probability of the Universal Decoder}
\label{upper}

The average probability of error of the proposed universal decoder,
is as follows.
\begin{eqnarray}
\bar{P}_{\mbox{\tiny e},q}&=&\sum_{\bx,\by} P_d(\bx,\by)\bar{P}_{\mbox{\tiny e},q}(\bx,\by)\\
&\dfn&\sum_{\bx,\by}P_d(\bx,\by)\mbox{Pr}\left\{(\hbX,\hbY)\ne (\bx,\by)\right\}\\
&=&\sum_{\bx,\by}P_d(\bx,\by)\cdot\mbox{Pr}\left[\bigcup_{\{(\bx',\by')\ne
(\bx,\by):~q(\bx',\by')\le
q(\bx,\by)\}}\left\{f(\bx')=f(\bx),~g(\by')=g(\by)\right\}\right].
\end{eqnarray}
As for $\bar{P}_{\mbox{\tiny e},q}(\bx,\by)$, we have
\begin{eqnarray}
\bar{P}_{\mbox{\tiny e},q}(\bx,\by)&=&\mbox{Pr}\left[\bigcup_{\{(\bx',\by')\ne
(\bx,\by):~q(\bx',\by')\le
q(\bx,\by)\}}\left\{f(\bx')=f(\bx),~g(\by')=g(\by)\right\}\right]\nonumber\\
&\le&\mbox{Pr}\left[\bigcup_{\{\bx'\ne\bx,~\by'\ne\by:~q(\bx',\by')\le
q(\bx,\by)\}}\left\{
f(\bx')=f(\bx),~g(\by')=g(\by)\right\}\right]+\nonumber\\
& &\mbox{Pr}\left[\bigcup_{\{\bx'\ne\bx:~q(\bx',\by)\le q(\bx,\by)\}}
\left\{f(\bx')=f(\bx)\right\}\right]+\nonumber\\
& &\mbox{Pr}\left[\bigcup_{\{\by'\ne\by:~q(\bx,\by')\le q(\bx,\by)\}}
\left\{g(\by')=g(\by)\right\}\right]\nonumber\\
&\dfn& \bar{P}_{\mbox{\tiny e},q,1}(\bx,\by)+
\bar{P}_{\mbox{\tiny e},q,2}(\bx,\by)+
\bar{P}_{\mbox{\tiny e},q,3}(\bx,\by).
\end{eqnarray}
Now,
\begin{eqnarray}
\bar{P}_{\mbox{\tiny
e},q,1}(\bx,\by)&=&\mbox{Pr}\left[\bigcup_{\{\tbx\ne\bx,~\tby\ne\by:~q(\tbx,\tby)\le
q(\bx,\by)\}}\left\{
f(\tbx)=f(\bx),~g(\tby)=g(\by)\right\}\right]\nonumber\\
&\le&\min\left\{1,2^{-n(R_x+R_y)}\bigg|\left\{(\tbx,\tby):~q(\tbx,\tby)\le
q(\bx,\by)\right\}\bigg|\right\}\nonumber\\
&=&\min\left\{1,2^{-n(R_x+R_y)}\bigg|\bigcup_{k=0}^{n-1}
\left\{(\tbx,\tby):~q_k(\tbx,\tby)\le
q(\bx,\by)\right\}\bigg|\right\}\nonumber\\
&\le&\min\left\{1,2^{-n(R_x+R_y)}\bigg|\bigcup_{k=0}^{n-1}
\left\{(\tbx,\tby):~q_k(\tbx,\tby)\le
q_d(\bx,\by)\right\}\bigg|\right\}\nonumber\\
&\le&\min\left\{1,2^{-n(R_x+R_y)}\sum_{k=0}^{n-1}
\bigg|\left\{(\tbx,\tby):~q_k(\tbx,\tby)\le
q_d(\bx,\by)\right\}\bigg|\right\}\nonumber\\
&\le&\min\bigg\{1,2^{-n(R_x+R_y)}\sum_{k=0}^{n-1}
\sum_{\{(\calT(\tx_1^{n-k},\ty_{k+1}^n),\calT(\ty_1^k),\calT(\tx_{n-k+1}^n):
~q_k(\tbx,\tby)\le
q_d(\bx,\by)\}}\nonumber\\
& &|\calT(\tx_1^{n-k},\ty_{k+1}^n)|\cdot
|\calT(\ty_1^k)|\cdot|\calT(\tx_{n-k+1}^n)|\bigg\}\nonumber\\
&\le&\min\bigg\{1,2^{-n(R_x+R_y)}\sum_{k=0}^{n-1}
\sum_{\{(\calT(\tx_1^{n-k},\ty_{k+1}^n),\calT(\ty_1^k),\calT(\tx_{n-k+1}^n):
~q_k(\tbx,\tby)\le
q_d(\bx,\by)\}}\nonumber\\
& &\exp_2\{(n-k)\hH(\tx_1^{n-k},\ty_{k+1}^n)+
k\hH(\ty_1^k)+k\hH(\tx_{n-k+1}^n)\}\bigg\}\nonumber\\
&=&\min\bigg\{1,2^{-n(R_x+R_y)}\sum_{k=0}^{n-1}\nonumber\\
& &\sum_{\{(\calT(\tx_1^{n-k},\ty_{k+1}^n),
\calT(\ty_1^k),\calT(\tx_{n-k+1}^n):~q_k(\tbx,\tby)\le
q_d(\bx,\by)\}}
2^{u_k(\tbx,\tby)}\bigg\}\nonumber\\
&\exe&\min\bigg\{1,\max_{0\le k\le n}\max_{\{\hP_{\ty_1^k},\hP_{\tx_1^{n-d}
\ty_{d+1}^n},\hP_{\tx_{n-d+1}^n}:~q_k(\tbx,\tby)\le q_d(\bx,\by)\}}\nonumber\\
& &\exp_2[u_k(\tbx,\tby)-n(R_x+R_y)]\bigg\}.
\end{eqnarray}
Similarly,
\begin{eqnarray}
\bar{P}_{\mbox{\tiny
e},q,2}(\bx,\by)&=&\mbox{Pr}\left[\bigcup_{\{\tbx\ne\bx:~q(\tbx,\by)\le q(\bx,\by)\}}
\left\{f(\tbx)=f(\bx)\right\}\right]\nonumber\\
&\le&\min\left\{1,2^{-nR_x}\bigg|\left\{\tbx:~q(\tbx,\by)\le
q(\bx,\by)\right\}\bigg|\right\}\nonumber\\
&\le&\min\left\{1,2^{-nR_x}\bigg|\bigcup_{k=0}^{n-1}\left\{\tbx:~q_k(\tbx,\by)\le
q_d(\bx,\by)\right\}\bigg|\right\}\nonumber\\
&\le&\min\left\{1,2^{-nR_x}\sum_{k=0}^{n-1}\bigg|\left\{\tbx:~q_k(\tbx,\by)\le
q_d(\bx,\by)\right\}\bigg|\right\}\nonumber\\
&\le&\min\bigg\{1,2^{-nR_x}\sum_{k=0}^{n-1}
\sum_{\{(\calT(\tx_1^{n-k}|y_{k+1}^n),\calT(\tx_{n-k+1}^n):~q_k(\tbx,\by)\le
q_d(\bx,\by)\}}\nonumber\\
& &|\calT(\tx_1^{n-k}|y_{k+1}^n)|\cdot|\calT(\tx_{n-k+1}^n)|\bigg\}\nonumber\\
&\le&\min\bigg\{1,2^{-nR_x}\sum_{k=0}^{n-1}
\sum_{\{(\calT(\tx_1^{n-k}|y_{k+1}^n),\calT(\tx_{n-k+1}^n):~q_k(\tbx,\by)\le
q_d(\bx,\by)\}}\nonumber\\
& &\exp_2\{(n-k)\hH(\tx_1^{n-k}|y_{k+1}^n)+k\hH(\tx_{n-k+1}^n)\}\bigg\}\nonumber\\
&=&\min\bigg\{1,2^{-nR_x}\sum_{k=0}^{n-1}
\sum_{\{(\calT(\tx_1^{n-k}|y_{k+1}^n),\calT(\tx_{n-k+1}^n):~q_k(\tbx,\by)\le
q_d(\bx,\by)\}}
2^{v_k(\tbx,\by)}\bigg\}\nonumber\\
&\exe&\min\left\{1,\max_{0\le k\le n}
\max_{\{\hP_{\tx_1^{n-k},y_{k+1}^n},\hP_{\tx_{n-k+1}^n}:~q_k(\tbx,\by)\le q_d(\bx,\by)\}}
\exp_2[v_k(\tbx,\by)-nR_x]\right\}\nonumber\\
&\le&\min\bigg\{1,\max_{0\le k\le n}\max_{\{\hP_{\ty_1^k},\hP_{\tx_1^{n-d}
\ty_{d+1}^n},\hP_{\tx_{n-d+1}^n}:~q_k(\tbx,\tby)\le q_k(\bx,\by)\}}\nonumber\\
& &\exp_2[v_k(\tbx,\tby)-nR_x]\bigg\},
\end{eqnarray}
and in exactly the same manner,
\begin{equation}
\bar{P}_{\mbox{\tiny
e},q,3}(\bx,\by)\le 
\min\left\{1,\max_{0\le k\le n}\max_{\{\hP_{\ty_1^k},\hP_{\tx_1^{n-d}
\ty_{d+1}^n},\hP_{\tx_{n-d+1}^n}:~q_k(\tbx,\tby)\le 
q_d(\bx,\by)\}}\exp_2[w_k(\tbx,\tby)-nR_y]\right\}.
\end{equation}
We now use the following simple inequality that holds for every non--negative
reals, $a$, $b$, and $c$:
\begin{eqnarray}
\min\{1,a\}+\min\{1,b\}+\min\{1,c\}&\le&3
\max\left\{\min\{1,a\},\min\{1,b\},\min\{1,c\}\right\}\\
&=&\left\{\begin{array}{ll}
3 & a> 1~\mbox{or}~b>1~\mbox{or}~c>1\\
3\cdot\max\{a,b,c\} & a\le 1~\mbox{and}~b\le 1~\mbox{and}~c\le 1\end{array}\right.\\
&=&\left\{\begin{array}{ll}
3 &\max\{a,b,c\}>1\\
3\cdot\max\{a,b,c\} & \max\{a,b,c\}\le 1\end{array}\right.\\
&=&3\cdot\min\left\{1,\max\{a,b,c\}\right\}.
\end{eqnarray}
It follows that
\begin{eqnarray}
\bar{P}_{\mbox{\tiny
e},q}(\bx,\by)&\lexe&3\cdot\min\bigg\{1,\max_{0\le k\le n}\max_{\{\hP_{\ty_1^k},\hP_{\tx_1^{n-d}
\ty_{d+1}^n},\hP_{\tx_{n-d+1}^n}:~q_k(\tbx,\tby)\le q_d(\bx,\by)\}}\nonumber\\
& &\exp_2\{\max\{u_k(\tbx,\tby)-n(R_x+R_y),v_k(\tbx,\tby)-nR_x,w_k(\tbx,\tby)-nR_y\}\}\bigg\}\\
&\exe&\min\left\{1,\max_{0\le k\le n}\max_{\{\hP_{\ty_1^k},\hP_{\tx_1^{n-d}
\ty_{d+1}^n},\hP_{\tx_{n-d+1}^n}:~q_k(\tbx,\tby)\le q_d(\bx,\by)\}}2^{q_k(\tbx,\tby}\right\}\\
&\le&\min\left\{1,2^{q_d(\bx,\by)}\right\}.
\end{eqnarray}
Finally, the overall average error probability, associated with the proposed universal
decoder, is exponentially upper bounded by
\begin{equation}
\label{ub}
\bar{P}_{\mbox{\tiny e},q}\lexe \bE\min\left\{1,\exp_2[q_d(\bX,\bY)]\right\}.
\end{equation}

\subsection{Lower Bound on the Error Probability of the MAP Decoder}
\label{lower}

For the MAP decoder, the conditional average probability of error, for a given $(\bx,\by)$,
is as follows:
\begin{eqnarray}
\bar{P}_{\mbox{\tiny e},\star}(\bx,\by)
&=&\mbox{Pr}\left[\bigcup_{\{(\tbx,\tby)\ne
(\bx,\by):~P_d(\tbx,\tby)\ge
P_d(\bx,\by)\}}\left\{f(\tbx)=f(\bx),~g(\tby)=g(\by)\right\}\right]\nonumber\\
&\ge&\mbox{Pr}\left[\bigcup_{\{(\tbx,\tby)\in\calS_{\mbox{\tiny o}}(\bx,\by)\}}\left\{
f(\tbx)=f(\bx),~g(\tby)=g(\by)\right\}\right],
\end{eqnarray}
where
\begin{equation}
\calS_{\mbox{\tiny o}}(\bx,\by)=\{(\tbx,\tby)\ne(\bx,\by):~\ty_1^d\in\calT(y_1^d),~
(\tx_1^{n-d},\ty_{d+1}^n)\in\calT(x_1^{n-d},y_{d+1}^n),
\tx_{n-d+1}^n\in\calT(x_{n-d+1}^n)\}.
\end{equation}
To further lower bound the conditional average probability of error, associated with
the MAP decoder, we need the following lemma, whose proof is deferred to the appendix.

\begin{lemma}
Let $\calS_{\mbox{\tiny o}}$ be a set of pairs of 
integers, $\{(i,j)\}$, with the
following properties. 
\begin{enumerate}
\item The pair $(0,0)$ is a not member of $\calS_{\mbox{\tiny o}}$.
\item For a given $i$,
let $\calS_{i*}=\{j:~(i,j)\in\calS_{\mbox{\tiny o}}\}$ and
$\calS_{*j}=\{i:~(i,j)\in\calS_{\mbox{\tiny o}}\}$. We assume that
$|\calS_{i*}|=\ell$ for all $i$
such that $(i,j)\in\calS_{\mbox{\tiny o}}$ for some $j$. Likewise,
$|\calS_{*j}|=k$ for all $j$ such that $(i,j)\in\calS_{\mbox{\tiny o}}$ for
some $i$. Here, $\ell$ and $k$ are fixed positive integers.
\item For every $(i,j)\in\calS_{\mbox{\tiny o}}$, there is an event $\calC_{ij}$, 
defined as $\calC_{ij}=\calA_i\cap \calB_j$, 
where $\{\calA_i\}$ and $\{\calB_j\}$ are sequences of mutually independent events.  
\item The probabilities of $\{\calA_i\}$ are given by $P[\calA_0]=1$ and
$P[\calA_i]=\alpha$ for all $i\ne 0$. Here, $\alpha$ is a fixed number in $[0,1]$.
\item The probabilities of $\{\calB_j\}$ are given by $P[\calB_0]=1$ and
$P[\calB_j]=\beta$ for all $j\ne 0$. Here, $\beta$ is a fixed number in $[0,1]$.
\end{enumerate}
Then, under conditions 1--5, 
\begin{equation}
P\left[\bigcup_{(i,j)\in\calS_{\mbox{\tiny o}}}\calC_{ij}\right]\ge
\frac{1}{4}
\cdot\min\left\{1,\max\left\{k\alpha,\ell\beta,(M-k-\ell)
\alpha\beta\right\}\right\},
\end{equation}
where $M=|\calS_{\mbox{\tiny o}}|$.
\end{lemma}

We apply Lemma~1 using the following assignments:
$i=\tbx$, $j=\tby$, and so, $i=0$ and $j=0$ correspond to
$\tbx=\bx$ and $\tby=\by$, respectively. The event $\calA_i$ is
$\{f(\tbx)=f(\bx)\}$, where $\bx$ and $f(\bx)$ are given. Thus, obviously
$P[\calA_0]=1$. Likewise, the event $\calB_j$ is $\{g(\tby)=g(\by)\}$ for a given
$\by$ and $g(\by)$ and $P[\calB_0]=1$. It follows then that
$\alpha=2^{-nR_x}$,
$\beta=2^{-nR_y}$, and $\calS_{\mbox{\tiny
o}}=\calS_{\mbox{\tiny o}}(\bx,\by)$.
Here, 
\begin{eqnarray}
M&\exe&\exp_2\{(n-d)\hH(x_1^{n-d},y_{d+1}^n)+
d\hH(x_{n-d+1}^n)+d\hH(y_1^d)\}=2^{u_d(\bx,\by)}\\
k&\exe&\exp_2\{(n-d)\hH(x_1^{n-d}|y_{d+1}^n)+d\hH(x_{n-d+1}^n)\}=2^{v_d(\bx,\by)}\\
\ell&\exe&\exp_2\{d\hH(y_1^d)+(n-d)\hH(y_{d+1}^n|x_1^{n-d})\}=2^{w_d(\bx,\by)}.
\end{eqnarray}
Thus, according to Lemma~1, we obtain the matching lower bound,
\begin{eqnarray}
\bar{P}_{\mbox{\tiny e
},\star}(\bx,\by)&\gexe&\min\left\{1,\max\left\{2^{-nR_x}\cdot
2^{v_d(\bx,\by)},2^{-nR_y}\cdot 2^{w_d(\bx,\by)},2^{-n(R_x+R_y)}\cdot
2^{u_d(\bx,\by)}\right\}\right\}\\
&\exe&\min\left\{1,\exp_2[q_d(\bx,\by)]\right\},
\end{eqnarray}
and so, the overall average error probability of the MAP decoder is exponentially
lower bounded by
\begin{equation}
\bar{P}_{\mbox{\tiny e},\star}\gexe\bE\min\left\{1,\exp_2[q_d(\bX,\bY)]\right\},
\end{equation}
which matches the upper bound of the universal decoder 
in eq.\ (\ref{ub}), as far as the exponential order goes.

\subsection{The Error Exponent Formula}
\label{formula}

From the previous subsections, we learn that both decoders have an average error
probability of the exponential order of $\bE\min\{1,\exp_2\{q_d(\bX,\bY)\}$.
Standard analysis of this quantity, using the well known method of types
\cite{CK81}, yields the
following single--letter expression: 
\begin{equation}
E_*(R_x,R_y)=E_q(R_x,R_y)=\min\{E_{xy}(R_x,R_y),E_{x|y}(R_x),E_{y|x}(R_y)\}, 
\end{equation}
where
\begin{eqnarray}
E_{xy}(R_x,R_y)&=&\min_{Q_{X'},Q_{Y'},Q_{XY}}\bigg\{\delta D(Q_{X'}\|P_X)+\delta
D(Q_{Y'}\|P_Y)+(1-\delta)D(Q_{X'Y'}\|P_{XY})+\nonumber\\
& &[R_x+R_y-\delta H_Q(X')-\delta
H_Q(Y')-(1-\delta)H_Q(X,Y)]_+\bigg\}\nonumber\\
E_{x|y}(R_x)&=&\min_{Q_{X'},Q_{XY}}\bigg\{\delta
D(Q_{X'}\|P_X)+(1-\delta)D(Q_{X'Y'}\|P_{XY})+\nonumber\\
& &[R_x-\delta
H_Q(X')-(1-\delta)H_Q(X|Y)]_+\bigg\}\\
E_{y|x}(R_x)&=&\min_{Q_{Y'},Q_{XY}}\bigg\{\delta
D(Q_{Y'}\|P_Y)+(1-\delta)D(Q_{X'Y'}\|P_{XY})+\nonumber\\
& &[R_y-\delta
H_Q(Y')-(1-\delta)H_Q(Y|X)]_+\bigg\}.
\end{eqnarray}
To find the Lagrange--dual of $E_{xy}(R_x,R_y)$, we proceed as follows.
\begin{eqnarray}
E_{xy}(R_x,R_y)&=&\min_{Q_{X'},Q_{Y'},Q_{XY}}\bigg\{\delta D(Q_{X'}\|P_X)+\delta
D(Q_{Y'}\|P_Y)+(1-\delta)D(Q_{X'Y'}\|P_{XY})+\nonumber\\
& &[R_x+R_y-\delta H_Q(X')-\delta
H_Q(Y')-(1-\delta)H_Q(X,Y)]_+\bigg\}\nonumber\\
&=&\min_{Q_{X'},Q_{Y'},Q_{XY}}\max_{0\le\rho\le 1}\bigg\{\delta
D(Q_{X'}\|P_X)+\delta
D(Q_{Y'}\|P_Y)+(1-\delta)D(Q_{X'Y'}\|P_{XY})+\nonumber\\
& &\rho[R_x+R_y-\delta H_Q(X')-\delta
H_Q(Y')-(1-\delta)H_Q(X,Y)]\bigg\}\nonumber\\
&\eqa&\max_{0\le\rho\le 1}\bigg[\rho(R_x+R_y)+\min_{Q_{X'},Q_{Y'},Q_{XY}}\bigg\{\delta
D(Q_{X'}\|P_X)+\nonumber\\
& &\delta D(Q_{Y'}\|P_Y)+(1-\delta)D(Q_{X'Y'}\|P_{XY})+\nonumber\\
& &\rho[R_x+R_y-\delta H_Q(X')-\delta
H_Q(Y')-(1-\delta)H_Q(X,Y)]\bigg\}\bigg]\nonumber\\
&=&\max_{0\le\rho\le 1}\bigg[\rho(R_x+R_y)+\bigg\{\delta\cdot\min_{Q_{X'}}[D(Q_{X'}\|P_X)-\rho
H_Q(X')]+\nonumber\\
& &\delta\cdot\min_{Q_{Y'}}[D(Q_{Y'}\|P_Y)-\rho H_Q(Y')]+\nonumber\\
& &(1-\delta)\cdot\min_{Q_{XY}}[D(Q_{XY}\|P_{XY})-\rho
H_Q(X,Y)]\bigg\}\bigg]\nonumber\\
&\eqb&\max_{0\le\rho\le 1}\bigg\{\rho(R_x+R_y)-\delta
\log\left(\sum_x
P_X(x)^{1/(1+\rho)}\right)^{1+\rho}-\nonumber\\
& &\delta\log\left(\sum_y
P_Y(y)^{1/(1+\rho)}\right)^{1+\rho}-
(1-\delta)\log\left(\sum_{x,y}
P_{XY}(x,y)^{1/(1+\rho)}\right)^{1+\rho}\bigg\}\nonumber\\
&=&\max_{0\le\rho\le 1}\rho\cdot\bigg[R_x+R_y-\delta
H_{1/(1+\rho)}(X)-\nonumber\\
& &\delta H_{1/(1+\rho)}(Y)-(1-\delta)H_{1/(1+\rho)}(X,Y)\bigg],
\end{eqnarray}
where in (a) we invoked the minimax theorem for convex--concave functions, and
in (b) we carried out the minimizations using standard methods.
The Lagrange--duals of $E_{x|y}(R_x)$ and $E_{y|x}(R_y)$ are obtained in a
similar fashion.

Finally, for an error exponent level, $E$, to be achievable by the random
code, $E_{xy}(R_x,R_y)$, $E_{x|y}(R_x)$ and $E_{y|x}(R_y)$ must all be at
least as large as $E$ at the same time. The condition $E_{xy}(R_x,R_y)\ge E$
is equivalent to the condition 
\begin{equation}
\exists~0\le\rho\le 1~~\rho\left[(R_x+R_y)-\delta H_{1/(1+\rho)}(X)-
\delta H_{1/(1+\rho)}(Y)-(1-\delta)H_{1/(1+\rho)}(X,Y)\right]\ge E
\end{equation}
or, equivalently,
\begin{equation}
\exists~0\le\rho\le 1~~R_x+R_y\ge \frac{E}{\rho}+\delta H_{1/(1+\rho)}(X)+
\delta H_{1/(1+\rho)}(Y)+(1-\delta)H_{1/(1+\rho)}(X,Y),
\end{equation}
which is the same\footnote{Change the variable $\rho$ to $s=1/\rho$.} as
\begin{equation}
\exists~s\ge 1~~R_x+R_y\ge sE+\delta H_{s/(1+s)}(X)+
\delta H_{s/(1+s)}(Y)+(1-\delta)H_{s/(1+s)}(X,Y),
\end{equation}
or
\begin{equation}
R_x+R_y\ge \inf_{s\ge 1}\left[sE+\delta H_{s/(1+s)}(X)+
\delta H_{s/(1+s)}(Y)+(1-\delta)H_{s/(1+s)}(X,Y)\right].
\end{equation}
Similarly, the requirements that $E_{x|y}(R_x)\ge E$
and $E_{y|x}(R_y)\ge E$ yield the individual lower bounds on $R_x$ and $R_y$,
that together form the achievable rate region, $\bR(E)$, as defined.


\section*{Appendix - Proof of Lemma 1}
\renewcommand{\theequation}{A.\arabic{equation}}
    \setcounter{equation}{0}

Let us partition $\calS_{\mbox{\tiny o}}$ into three disjoint subsets whose
union is equal to $\calS_{\mbox{\tiny o}}$. These subsets are
\begin{eqnarray}
\calS_{0*}&=&\{(0,j)\in\calS_{\mbox{\tiny o}}\}\\
\calS_{*0}&=&\{(i,0)\in\calS_{\mbox{\tiny o}}\}\\
\calS&=&\{(i,j)\in\calS_{\mbox{\tiny o}}:~i\ne 0,~j\ne 0\},
\end{eqnarray}
Clearly,
\begin{equation}
P\left[\bigcup_{(i,j)\in\calS_{\mbox{\tiny o}}}\calC_{ij}\right]\ge
\max\left\{P\left[\bigcup_{(i,j)\in\calS_{*0}}\calC_{ij}\right],
P\left[\bigcup_{(i,j)\in\calS_{0*}}\calC_{ij}\right],
P\left[\bigcup_{(i,j)\in\calS}\calC_{ij}\right]\right\}.
\end{equation}
As for the first term on the r.h.s., we have
\begin{eqnarray}
P\left[\bigcup_{(i,j)\in\calS_{*0}}\calC_{ij}\right]&=&
P\left[\bigcup_i \calA_i\cap \calB_0\right]\\
&=&P\left[\calB_0\bigcap\left(\bigcup_i \calA_i\right)\right]\\
&=&P[\calB_0]\cdot P\left[\bigcup_i \calA_i\right]\\
&=&P\left[\bigcup_i \calA_i\right]\\
&\ge&\frac{1}{2}\cdot\min\{1,k\alpha\},
\end{eqnarray}
where the last step follows
from de Caen's lower bound \cite{decaen97} on the probability 
of a union of a finite set of events,
$\{\calE_i,~i\in\calI\}$: 
\begin{equation} 
P\left[\bigcup_{i\in\calI}\calE_i\right]\ge
\sum_{i\in\calI} \frac{P^2[\calE_i]}{\sum_{j\in\calI}P[\calE_i\cap\calE_j]},
\end{equation}
which in the case of pairwise independent events, simplifies to
\begin{eqnarray}
P\left[\bigcup_{i\in\calI}\calE_i\right]&\ge&\sum_{i\in\calI}
\frac{P^2[\calE_i]}{P[\calE_i]+\sum_{j\in\calI}P[\calE_i]\cdot P[\calE_j]}\nonumber\\
&=&\sum_{i\in\calI}
\frac{P[\calE_i]}{1+\sum_{j\in\calI}P[\calE_j]}\nonumber\\
&=&\frac{\sum_{i\in\calI}P[\calE_i]}{1+\sum_{i\in\calI}P[\calE_i]}\nonumber\\
&\ge&\frac{\sum_{i\in\calI}
P[\calE_i]}{2\cdot\max\{1,\sum_{i\in\calI}P[\calE_i]\}}\nonumber\\
&=&\frac{1}{2}\cdot\min\left\{1,\sum_{i\in\calI}P[\calE_i]\right\}.
\end{eqnarray}
Similarly,
\begin{equation}
P\left[\bigcup_{(i,j)\in\calS_{*0}}\calC_{ij}\right]\ge\frac{1}{2}\cdot\min\{1,\ell\beta\}.
\end{equation}

Moving on to the union over $\calS$,
let us denote $\calU=\{i:~\exists
j~(i,j)\in\calS\}$ and
$\calV=\{j:~\exists i~(i,j)\in\calS\}$, with
cardinalities $K$ and $L$, respectively. Then, by property 2 of $\calS_{\mbox{\tiny o}}$,
$M'\dfn|\calS|=K\ell
=kL$.\footnote{Note that in general, $M'\le
KL$. For example, if $\calS=\{(1,1), (2,1),
(2,2), (3,2), (3,3), (1,3)\}$, then $M'=|\calS|=6$, $k=\ell=2$,
and $\calU=\calV=\{1,2,3\}$, so $K=L=3$.
As another example: let $(i,j)$ designate
indexes of finite--alphabet $n$-sequences
and let $\calS$ be a joint type, $\calT(i,j)$, then $\calU=\calT(i)$,
$\calS_i=\calT(j|i)$, $\calV=\calT(j)$ and $\calS_j=\calT(i|j)$, but
$\calT(i,j)$ is, in general, only a subset of $\calT(i)\times\calT(j)$.}
Applying again de Caen's lower bound, this time, to the union over $\calS$, we have
\begin{eqnarray}
P\left[\bigcup_{(i,j)\in\calS}
\calC_{ij}\right]&\ge&
\sum_{(i,j)\in\calS}\frac{P^2(\calC_{ij})}
{P(\calC_{ij})+\sum_{(i',j')\in\calS\setminus\{(i,j)\}}
P(\calC_{ij}\cap \calC_{i'j'})}\nonumber\\
&=&\sum_{(i,j)\in\calS}P^2(\calA_i)P^2(\calB_j)\cdot\bigg(P(\calA_i)P(\calB_j)
+\sum_{i'\in\calS_j\setminus\{i\}}
P[\calA_i\cap \calA_{i'}\cap \calB_j]+\nonumber\\
& &\sum_{j'\in\calS_i\setminus\{j\}}P[\calA_i\cap\calB_j\cap\calB_{j'}]+
\sum_{\{(i',j')\in\calS:~i'\ne i,~j'\ne j\}}P[\calA_i\cap\calA_{i'}\cap\calB_j\cap
\calB_{j'}]\bigg)^{-1}\nonumber\\
&\ge&\sum_{(i,j)\in\calS}\frac{\alpha^2\beta^2}{\alpha\beta+k\alpha^2\beta+
\ell\alpha\beta^2+M'\alpha^2\beta^2}\nonumber\\
&=&\sum_{(i,j)\in\calS}\frac{\alpha\beta}{1+k\alpha+
\ell\beta+M'\alpha\beta}\nonumber\\
&=&\frac{M'\alpha\beta}{1+k\alpha+
\ell\beta+M'\alpha\beta}\nonumber\\
&\ge&\frac{M'\alpha\beta}{4\cdot\max\{1,k\alpha,
\ell\beta,M'\alpha\beta\}}\nonumber\\
&=&\frac{1}{4}\cdot\min\left\{1,\frac{M'\alpha}{\ell},
\frac{M'\beta}{k},M'\alpha\beta\right\}\nonumber\\
&=&\frac{1}{4}\cdot\min\left\{1,K\alpha,L\beta,M'\alpha\beta\right\}.
\end{eqnarray}
Thus, overall we have
\begin{equation}
P\left[\bigcup_{(i,j)\in\calS_{\mbox{\tiny o}}}
\calC_{ij}\right]\ge\frac{1}{4}\cdot\max\left\{\min\{1,k\alpha\},\min\{1,\ell\beta\},
\min\{1,K\alpha,L\beta,M'\alpha\beta\}\right\}.
\end{equation}
Now, consider the following line of thought: if $M'\alpha\beta\ge
K\alpha$, which is equivalent to $\ell\beta\ge 1$, then the lower bound
is at least as large as $1/4$.
Similarly, if $M'\alpha\beta\ge L\beta$, which is
equivalent to $k\alpha\ge 1$, then again the lower bound is at least $1/4$. Thus,
\begin{eqnarray}
P\left[\bigcup_{(i,j)\in\calS_{\mbox{\tiny o}}}
\calC_{ij}\right]&\ge&\frac{1}{4}\cdot\left\{\begin{array}{ll}
1 & k\alpha\ge 1
~\mbox{or}~\ell\beta\ge 1\\
\max\{k\alpha,\ell\beta,\min\{1,M'\alpha\beta\}\} &
\mbox{otherwise}\end{array}\right.\\
&=&\frac{1}{4}\cdot\left\{\begin{array}{ll}
1 & k\alpha\ge 1~\mbox{or}~\ell\beta\ge 1~\mbox{or}~M'\alpha\beta\ge
1\\
\max\{k\alpha,\ell\beta,M'\alpha\beta\} &
\mbox{otherwise}\end{array}\right.\\
&=&\frac{1}{4}\cdot\left\{\begin{array}{ll}
1 & \max\{k\alpha,\ell\beta,M'\alpha\beta\}\ge
1\\
\max\{k\alpha,\ell\beta,M'\alpha\beta\} &
\max\{k\alpha,\ell\beta,M'\alpha\beta\}<1
\end{array}\right.\\
&=&\frac{1}{4}\cdot\min\left\{1,\max\{k\alpha,\ell\beta,M'\alpha\beta\}\right\},
\end{eqnarray}
and the proof of the lemma is completed upon observing that
$M'=M-k-\ell$.


\newpage
\begin{thebibliography}{AA}

\bibitem{CHJL08}
J.~Chen, D.-k.~He, A.~Jagmohan, and L.~A.~Lastras--Monta\~no, ``
On universal variable--rate Slepian--Wolf coding,''
{\it Proc.\ 2008 IEEE International Conference on Communications (ICC 2008)},
pp.\ 1426--1430, 2008.

\bibitem{Csiszar80}
I.~Csisz\'ar, ``Joint source–-channel error exponent,'' {\it Problems of Control 
and Information Theory}, vol.\ 9, no.\ 5, pp.\ 315–-328, 1980.

\bibitem{Csiszar82}
I.~Csisz\'ar, ``Linear codes for sources and source networks: error 
exponents, universal coding,''
{\it IEEE Trans.\ Inform.\ Theory}, vol.\ IT–-28, no.\ 4, pp.\ 585-–592, July 1982.

\bibitem{CK80}
I.~Csisz\'ar and J.~K\"orner, ``Towards a general 
theory of source networks,'' {\it IEEE Trans.\ Inform.\
Theory}, vol.\ IT--26, no.\ 2, pp.\ 155-165, March 1980.

\bibitem{CK81}
I.~Csisz\'ar and J.~K\"orner, {\it Information Theory: Coding Theorems 
for Discrete Memoryless
Systems}, Academic Press, 1981.
Second Edition: Cambridge University Press, New York, 2011.

\bibitem{decaen97}
D.~de Caen, ``A lower bound on the probability of a union,''
{\it Discrete Mathematics}, vol.\ 169, pp.\ 217--220, 1997.

\bibitem{Draper04}
S.~C.~Draper, ``Universal incremental Slepian--Wolf coding,'' {\it Proc.\ 42nd
Annual Allerton Conference on Communication, Control and Computing},
Monticello, IL, USA, October 2004.

\bibitem{Kieffer80}
J.~C.~Kieffer, ``Some universal noiseless multiterminal source coding
theorems,'' {\it Information and Control}, vol.\ 46, pp.\ 93--107, 1980.


\bibitem{TU13}
T.~Matsuta and T.~Uyematsu, ``Universal coding for asynchronous Slepian--Wolf coding
systems,'' {\it IEICE Tech.\ Rep.}, vol.\ 112, no.\ 460, pp.\ 1--6, March 2013.

\bibitem{TU15}
T.~Matsuta and T.~Uyematsu, ``Achievable rate regions for asynchronous Slepian--Wolf coding
systems,'' {\it Proc.\ 2015 IEEE Workshop on 
Information Theory}, pp.\ 312--316, October 2015.

\bibitem{TU20}
T.~Matsuta and T.~Uyematsu, ``Coding theorems for asynchronous Slepian--Wolf coding
systems,'' {\it IEEE Trans.\ Inform.\ Theory}, vol.\ 66, no.\ 8, pp.\ 
4774--4795, August 2020.

\bibitem{p183}
N.~Merhav, ``Universal decoding for source--channel coding with side
information,'' {\it Communications in Information and Systems},
vol.\ 16, no.\ 1, pp.\ 17--58, 2016.

\bibitem{p191}
N.~Merhav, ``Universal decoding using a noisy codebook,''
{\it IEEE Trans.\ Inform.\ Theory}, vol.\ 64, part 1, no.\ 4, pp.\ 2231--2239,
April 2018.

\bibitem{OH94}
Y.~Oohama and T.~S.~Han, ``Universal coding for
the Slepian--Wolf data compression system
and the strong converse theorem,'' {\it
IEEE Trans.\ Inform.\ Theory}, vol.\ 40, no.\ 6,
pp.\ 1908--1919, November 1994.

\bibitem{OO97}
N.~Oki and Y.~Oohama, ``Coding for the
asynchronous Slepian--Wolf data compression,''
{\it Proc.\ 20th Symposium on Information
Theory and its Applications (SITA `97)},
pp.\ 89--92, December 1997.


\bibitem{RU97}
B.~Rimoldi and R.~Urbanke, ``Asynchronous Slepian--Wolf coding via source--splitting,''
{\it Proc.\ 1997 IEEE International Symposium on Information Theory (ISIT `97)},
p.\ 271, Ulm, Germany, June 1997.

\bibitem{SBB05}
S.~Sarvotham, D.~Baron, and R.~G.~Baraniuk, ``Variable--rate universal
Slepian--Wolf coding with feedback,'' {\it Proc.\ 39th Asilomar Conference on 
Signals, Systems and Computers}, pp.\ 8--12, November 2005.

\bibitem{SW73}
D.~Slepian and J.~Wolf, ``Noiseless coding of correlated information sources,''
{\it IEEE Trans.\ Inform.\ Theory}, vol.\ IT-19, no.\ 4, pp.\ 471--480, July 1973.

\bibitem{STCW10}
Z.~Sun, C.~Tian, J.~Chen, and K.~M.~Wong, ``LDPC code design for asynchronous
Slepian--Wolf coding,'' {\it IEEE Trans.\ Commun.}, vol.\ 58, no.\ 2, pp.\ 511--520,
February 2010.

\bibitem{Willems88}
F.~M.~J.~Willems, ``Totally asynchronous Slepian--Wolf data compression,'' {\it IEEE
Trans.\ Inform.\ Theory}, vol.\ 34, no.\ 1, pp.\ 35--44, January 1988.

\bibitem{Ziv85}
J.~Ziv, ``Universal decoding for finite-state channels,'' 
{\em IEEE Trans.~Inform.~Theory\/},
vol.~IT--31, no.~4, pp.~453--460, July 1985.

\bibitem{ZL78}
J.~Ziv and A.~Lempel, ``Compression of individual sequences via 
variable-rate coding,''
{\em IEEE Trans.~Inform.~Theory\/},
vol.~IT--24, no.~5, pp.~530--536, September 1978.
\end{thebibliography}
\end{document}